\documentclass{article}
\usepackage{epsfig}
\begin{document}
\title{Radiating fluid spheres in the\\ effective 
vari\-ables ap\-prox\-i\-ma\-tion}
\author{W. Barreto and 
B. Rodr\'\i guez\\
Grupo de Relatividad y Astrof\'\i sica Te\'orica,\\
           Departamento de F\'\i sica,
           Escuela de Ciencias, N\'ucleo de Sucre,\\
           Universidad de Oriente,
           Cuman\'a, Venezuela\\
H. Mart\'\i nez\\
Secci\'on de F\'\i sica,Unidad de Estudios Generales,\\
           Universidad Nacional Experimental 
           Polit\'ecnica Antonio
           Jos\'e de Sucre,\\ Puerto Ordaz, Venezuela}
\maketitle
\begin{abstract}
We study the evolution
of spherically symmetric radiating fluid distributions
using the effective variables method, implemented {\it ab initio}
in Schwarzschild coordinates.
To illustrate the procedure and to establish some comparison with
 the original
method, we integrate numerically the set of
 equations at the surface
for two different models. 
The first model
is derived from the Schwarzschild interior solution. 
The second model is inspired in the Tolman VI solution.
\end{abstract}
\section{INTRODUCTION}
In 1980, Herrera, Jim\'enez and Ruggeri proposed a seminumerical
me\-thod, hereafter referred to as the effective variables method (EVM),
that can be used to
obtain nonstatic models from static solutions. This method
divides the spacetime in two spatial regions. The outer region
is described by the Vaidya solution and the spacetime metric
in the interior is obtained by solving the Einstein field
equations. Further, proper boundary conditions are imposed
in order to guarantee a smooth matching of both solutions.
The EVM has been used extensively to study astrophysical scenarios 
in radiative coordinates
\cite{hjr80,hn90}. The method assumes 
that the called effective
variables $\tilde\rho$ and $\tilde p$, which depend also on the
time--like coordinate, have the same radial dependence as the
corresponding static physical variables (energy density and pressure)
obtained from a static interior solution of the Einstein equations.
The rationale behind such an assumption is the fact that
 the effective variables
reduce to their physical counterparts in the static limit.
This approach can be justified by means of 
the characteristic times for different processes which take place
in the collapse scenario \cite{m96,kw90,bl86}.
If the hydrostatic time scale, $\tau_{hydr.}\approx 1/\sqrt{G\rho}$, 
is much shorter than the Kelvin--Helmholtz time scale $\tau_{KH}$, then 
in a first approximation the inertial terms in the equation of motion
can be ignored. Consequently, it seems reasonable to assume, in this
approximation, that the radial dependence of the physical variables
is the same as in the static solution. However, a better
approximation is obtained by assuming that the effective variables,
not the physical ones, have the same radial dependence
as the corresponding physical variables of the static situation \cite{hjr80}.

If the EVM is truly general,
an implementation in Schwarzschild coordinates 
 could be more
interesting for astrophysicists since these are the type of
coordinates commonly used by them; this 
is successfully accomplished in this paper. 
The idea that we can always construct dynamical solutions from static
ones seems a general method. Besides, as we will show, the EVM in
Schwarzschild coordinates introduces higher dynamic corrections, 
by means of the velocity respect to a minkowskian observer, than those
obtained in the Bondi coordinates treatment.

 In this paper we model two
simple but physically meaningful scenarios. The
first is obtained from a static solution for an incompressible fluid.
Once the sphere departs from static equilibrium by the emission of
energy, it slowly recovers the
initial state (staticity), unless some energy can be reabsorbed at the
surface to reach a constant radius in short time.
 In fact we show that even in Bondi coordinates the same result
 holds if staticity at the surface is enforced. The second model corresponds
to a highly compressed gas of fermions, Tolman's solution VI \cite{t39},
and leads us to an exploding sphere as has been reported 
\cite{hjr80,b93}. Additionally, these seminumerical models could serve
as useful test beds
for the numerical relativity methods being
 developed to match Cauchy and
characteristic codes \cite{pitt}--\cite{ddc95}

The paper is organized as follows. In section 2 we
write the field equations, for the inner region, in Schwarzschild
coordinates. Describing the
exterior space--time by means of the Vaidya metric, in section 3,
we treat the matching conditions and write the equations at the
boundary of the distribution of matter. In the section 4 we
show our two models; finally we conclude in section 5.

\section{FIELD EQUATIONS}
~~~~To write the Einstein field equations, inside the distribution
of matter, we use the line element in 
Schwarzschild coordinates
\begin{equation}
ds^2=e^\nu dt^2-e^\lambda dr^2-r^2\left( d\theta ^2+\sin
{}^2\theta\, d\phi ^2\right), \label{eq:metric} 
\end{equation}
where $\nu = \nu(t,r)$ and $\lambda = \lambda(t,r)$, with
 $(t,r,\theta,\phi)\equiv(0,1,2,3)$.

Physical input is obtained by introducing 
 Minkowski coordinates $(\tau,x,y,z)$ by \cite{b64}
\begin{equation}
d\tau=e^{\nu /2}dt,\,  
dx=e^{\lambda /2}dr,\,  
dy=rd\theta,\, 
dz=r \sin \theta d\phi.\label{eq:local}
\end{equation}
In these expressions $\nu$ and $\lambda$ are constants, because they
 only have local values. 

Next we assume that, for an observer moving relative to these coordinates
 with velocity $\omega$ in the radial ($x$) direction, the space is filled
with a fluid of density $\rho$, pressure $p$, and
unpolarized radiation of energy density $\hat \epsilon$.
For this comoving observer, the covariant energy tensor in Minkowski
 coordinates is 
 
\begin{equation}
\left(
\begin{array}{cccc}
\rho+\hat \epsilon & -\hat \epsilon & 0 & 0 \\
-\hat \epsilon & p +\hat \epsilon & 0 & 0 \\
0 & 0 & p & 0 \\
0 & 0 & 0 & p 
\end{array}
\right).
\end{equation}

Note that from (\ref{eq:local}) the velocity of matter in the Schwarzschild
 coordinates is
\begin{equation}
\frac{dr}{dt} = \omega e^{(\nu-\lambda)/2}. \label{eq:velocity}
\end{equation}

Now, by means of a Lorentz boost and defining 
$\epsilon \equiv \hat \epsilon(1+\omega)/(1-\omega)$ we write 
 the field equations in relativistic units ($G=c=1$) as follows: 
\begin{equation}
\frac{\rho + p \omega^2}{1-\omega ^2} + \epsilon =
\frac{1}{8\pi r}\left[\frac{1}{r} - 
e^{-\lambda}\left(\frac 1{r}-\lambda_{,r}\right)\right], \label{eq:ee1}
\end{equation}

\begin{equation}
\frac{p + \rho \omega^2}{1-\omega ^2} + \epsilon =
\frac{1}{8\pi r}\left[
e^{-\lambda}\left(\frac 1{r}+\nu_{,r}\right) - \frac{1}{r}\right], \label{eq:ee2}
\end{equation}

\begin{eqnarray}
p = \frac{1}{32\pi} \Big\{ e^{-\lambda}[ 2\nu_{,rr}+\nu_{,r}^2
-\lambda_{,r}\nu_{,r} + \frac{2}{r}
(\nu_{,r}-\lambda_{,r}) ] - \nonumber \\ \nonumber \\
e^{-\nu}[ 2\lambda _{,tt}+\lambda_{,t}(\lambda_{,t}-\nu_{,t}) ] \Big\},
 \label{eq:ee3}
\end{eqnarray}

\begin{equation}
(\rho + p)\frac{\omega}{1-\omega^2} + \epsilon = 
-\frac{\lambda_{,t}}{8\pi r}e^{-\frac 12(\nu+\lambda)}, \label{eq:ee4}
\end{equation}
where the comma subscript represents partial differentiation with 
 respect to the indicated
 coordinate.

We have four field equations for four physical variables ($\rho$, $p$,
 $\epsilon$ and $\omega$) and two geometrical
 variables ($\nu$ and $\lambda$). Obviously, additional information
is required 
 to handle the problem consistently. First, however, we 
 discuss the matching with the exterior solution and the surface
 equations that govern the dynamics.

\section{MATCHING CONDITIONS AND SURFACE EQUATIONS}
~~~~We describe the exterior space--time by the Vaidya metric 
\begin{equation}
ds^2=\left( 1-\frac{2{\cal M}(u)}R\right) du^2+2du\,dR-R^2\left( d\theta
^2+\sin^2\theta\, d\phi^2 \right),
\end{equation}
where $u$ is a time--like coordinate so that $u=$ constant represents,
 asymptotically,  null
 cones open to the future and $R$ is a null coordinate ($g_{RR}=0$). 

The exterior and interior solutions are separated by the surface $r=a(t)$.
 To match both regions on this surface we require the Darmois matching
 conditions. 
 Thus, demanding the continuity of the first fundamental form, we obtain
\begin{equation}
e^{-\lambda_a}=1-\frac{2{\cal M}}{R_a} \label{eq:ffa}
\end{equation}
and 
\begin{equation}
\nu_a = -\lambda_a. \label{eq:ffb}
\end{equation}
From now on, the subscript $a$ indicates that the quantity is evaluated 
 at the surface.
Matching conditions are usually obtained from the continuity of
the first and second fundamental forms. Here, however, we will use
the continuity of the independent components of the energy--momentum
flow instead of the second fundamental form, which have been shown
to be equivalent \cite{hd97} but it is simpler to apply in the
present case.
 This last condition
 guarantees absence of singular behaviors on the
 surface. It is easy to check that  
\begin{equation}
 p_a = 0, \label{eq:boundary}
\end{equation}
which expresses the continuity of the radial pressure.

To write the surface equations we introduce the mass function $m$
 by means of
\begin{equation}
e^{-\lambda (r,t)}=1-2 m(r,t)/r. \label{eq:masa}
\end{equation}
Substituting (\ref{eq:masa}) into (\ref{eq:ee1}) and (\ref{eq:ee4})
 we obtain, after some rearrangements, 
\begin{equation}
\frac{dm}{dt}=-4\pi r^2\left[\frac{dr}{dt}p
+\epsilon (1-\omega )\Big(1-\frac{2m}{r}\Big)^{1/2}e^{\nu /2} \right]. \label{eq:energy}
\end{equation}
This equation shows the energetics across the moving boundary of the 
fluid sphere. Evaluating (\ref{eq:energy}) at the surface and using the boundary
 condition (\ref{eq:boundary}), the
 energy loss is given by   
\begin{equation}
\dot m_a =-4\pi a^2 \epsilon _a (1 - 2m_a/a) (1-\omega_a).
\end{equation}
Hereafter, a dot over any variable indicates $d/dt$.
 The evolution of the boundary is governed by equation (\ref{eq:velocity})
 evaluated at the surface
\begin{equation}
\dot a=(1-2m_a/a)\omega _a.
\end{equation}
Scaling the total mass $m_a$, the radius $a$ and
 the time--like coordinate by the initial mass $m_a(t=0)\equiv m_a(0)$,
$$ A\equiv a/m_a(0), \, M\equiv m_a/m_a(0), \, t/m_a(0) \rightarrow t,$$
it is convenient to define
\begin{equation}
F\equiv 1-2M/A,
\end{equation}
\begin{equation}
\Omega \equiv \omega _a.
\end{equation}

Also we define the luminosity as seen by a comoving observer as \cite{m96}
\begin{equation}
\hat E\equiv(4\pi r^2\hat\epsilon)_{r=a},
\end{equation}
and the luminosity perceived by an observer at rest at infinity as
\begin{equation}
L\equiv-\dot M=F \hat E (1+\Omega).
\end{equation}
The function $F$ is related to the boundary redshift $z_a$ by
\begin{equation}
1+z_a=\frac{\nu_{em}}{\nu_{rec}}=F^{-1/2}.
\end{equation}
Thus the luminosity as measured by a noncomoving observer located
on the surface is
\begin{equation}
E=L(1+z_a)^2=-\frac{\dot M}{F}=\hat E (1+\Omega),
\end{equation}
where the term $(1+\Omega)$ accounts for the boundary Doppler shift.
With these definitions 
the surface equations can be written as
\begin{equation}
\dot A=F\Omega, \label{eq:first}
\end{equation}
\begin{equation}
\dot F=\frac{(1-F)\dot A +2L}{A}. \label{eq:second}
\end{equation}
Equations (\ref{eq:first}) and (\ref{eq:second}) are general
 within spherical symmetry. We need a third surface equation to specify
 the dynamics completely for any set of initial conditions and a given
 luminosity profile $L(t)$. For this purpose we can use 
 the conservation equation $T_{1;\mu }^\mu=0$ evaluated at the
 surface.  After straightforward manipulations the condition
 $T_{1;\mu }^\mu=0$ results in
\begin{eqnarray}
\tilde p_{,r} + \frac{(\tilde\rho + \tilde p)(4\pi r^3\tilde p + m)}{r(r-2m)} 
=\nonumber \\ \nonumber \\\frac{e^{-\nu}}{4\pi r(r-2m)}\left( m_{,tt} +\frac{3m_{,t}^2}{r-2m}-
\frac{m_{,t}\nu_{,t}}{2}\right) +
 \frac{2}{r}(p-\tilde p),\label{eq:TOV}
\end{eqnarray}
where the effective variables are defined by
\begin{equation}
\tilde \rho \equiv \frac{\rho + p \omega^2}{1-\omega ^2} + \epsilon
\end{equation}
and
\begin{equation}
\tilde p \equiv \frac{p + \rho \omega^2}{1-\omega ^2} + \epsilon.
\end{equation}
These effective variables are essentially the same as have been
defined by Herrera and colaborators, but now the velocity $\omega$
introduces a higher dynamics correction (quadratic). This fact could
be of interest to investigate its effect on dissipative processes
(like heat flow and viscosity).

Equation (\ref{eq:TOV}) is the generalization of the 
Tolman--Oppenheimer--Volkov equation for hydrostatic support in
 nonstatic radiative situations. Our equation 
 leads at the surface to a differential equation for
 $\Omega$ if we specify in some
 way the geometrical variables.
\section{MODELING}
From (\ref{eq:ee1}), (\ref{eq:ee2}) and (\ref{eq:masa}), easily we obtain
\begin{equation}
m=\int^{r}_{0}4\pi r^2 \tilde \rho dr, \label{eq:mass}
\end{equation}
\begin{equation}
\nu=\nu_{a} + \int^r_a \frac{2(4\pi r^3 \tilde p + m)}{r(r-2m)}dr. \label{eq:nu}
\end{equation}
Thus, $m$ and $\nu$ are expressed in terms of $\tilde \rho$ and $\tilde p$
 in the nonstatic case in the same way they are in terms of $\rho$ and $p$
 in the static case. These considerations suggest the application of
 the EVM which until now has been exclusively used in Bondi coordinates
 \cite{hjr80,hn90}; that is, we assume that the $r$ dependence
 on $\tilde \rho$
 and $\tilde p$ is the same as on the $\rho_{static}$ and $p_{static}$.

To illustrate the procedure, in what follows we model two simple 
scenarios which correspond to an
 incompressible fluid and to a highly compressed gas of fermions.

 \subsection{Schwarzschild--like model}
 Consider the well known 
 Schwarzschild interior solution, where the density
satisfies $\rho=constant$. Thus, in the EVM we take the effective 
density as
\begin{equation}
\tilde \rho = 
f(t),
\end{equation}
where $f$ is an arbitrary function of $t$. Now, with $\rho=constant$
we can integrate equation (\ref{eq:TOV}) in the static case and obtain
the expression for $p$ which leads us to
\begin{equation}
\frac{\tilde p + \frac{1}{3}\tilde\rho}{\tilde p + \tilde\rho}=
(1-\frac{8\pi}{3}\tilde\rho r^2)^{1/2}k(t), \label{eq:prep}
\end{equation}
where $k$ is a function of $t$ to be defined from the boundary condition
(\ref{eq:boundary}), which now reads, in terms of the effective variables, as
\begin{equation}
\tilde p_a=\tilde\rho_a\Omega^2 + \epsilon_a(1-\Omega^2). \label{eq:boun}
\end{equation}
Thus, (\ref{eq:prep}) and (\ref{eq:boun}) give
\begin{equation}
\tilde\rho=\frac{3(1-F)}{8\pi a^2},
\end{equation}
\begin{equation}
\tilde p=\frac{\tilde\rho}{3}\Biggl\{\frac{\chi_S\sqrt F -3\psi_S\xi}
{\psi_S\xi -\chi_S\sqrt F}\Biggr\}, \label{eq:effepre}
\end{equation}
with
$$
\xi=[1-(1-F)(r/a)^2]^{1/2} 
$$
and
$$\chi_S=3(\Omega^2+1)(1-F)+2E(1+\Omega),$$

$$\psi_S=(3\Omega^2+1)(1-F)+2E(1+\Omega).$$

Using (\ref{eq:mass}) and (\ref{eq:nu}) it is easy to obtain expressions
for $m$ and $\nu$:
\begin{equation}
m=m_a(r/a)^3, \label{eq:mass_sch}
\end{equation}
\begin{equation}
e^{\nu}=\Biggl\{\frac{\chi_S\sqrt F-\psi_S\xi}{2(1-F)}
\Biggr\}^2. \label{eq:nu_sch}
\end{equation}
In order to write down explicitely the surface equations for this example,
 it is interesting to note that the left side of (\ref{eq:TOV})
 is zero for any value
 of $f(t)$. Next, evaluating (\ref{eq:TOV}) at the surface, we obtain
\begin{eqnarray}
\dot\Omega=\frac{2}{3F(1-F)}\Bigg\{\Big[\frac{3}{AF}
\Big(FE+\frac{3}{2}(1-F)\dot A\Big)\nonumber\\ \nonumber \\
+\Big(\dot F-\frac{\dot A\psi_S}{A}\Big)\Big]\Big(FE+\frac{3}{2}(1-F)\dot A\Big)
\nonumber \\ \nonumber \\
 + \frac{6FE\dot A}{A}+ \frac{6(1-F)\dot A^2}{A}-\dot F E -F\dot E
\nonumber\\ \nonumber \\-\frac{3}{2}(1-F)\dot F\Omega
-\frac{F^2}{A}[\psi_S-(1-F)]\Bigg\}.
\end{eqnarray}
This last equation, together with (\ref{eq:first}) and (\ref{eq:second}),
 constitute
 the differential system for the surface in this example.  It is
necessary to specify one function of $t$ and the initial data. To compare
with Ref. \cite{hjr80} we choose $L$
 to be a gaussian and radiating away $1/10$ of
the initial mass. 
Therefore, the system
 can be numerically integrated for the following
 initial conditions (among others):
$$
A(0)=5.0, \,\, F(0)=0.6, \,\, \Omega=0.0.
$$
 The integration was done
 up to some $t$ with
 good behavior in the physical variables. 

Feeding back the numerical values of $A$, $F$ and $\Omega$ (and their
 derivatives)
 in (\ref{eq:mass_sch}) and (\ref{eq:nu_sch}) we obtain $m$ and $\nu$ 
(and their partial derivatives) for any
 value of $r$. Thus, functions $\rho$, $p$, $dr/dt$ and $\epsilon$ can be
 monitored for any piece of the material, via field equations.
 We calculated
 them for the values $r/a= 0.2,\,\, 0.4,\,\, 0.6,\,\, 0.8$ and
 $1.0$.   It is interesting to observe
that once the sphere departs from static equilibrium by means of an
emission of energy, it will not recover a constant radius at all,
 at least within the
integrated interval of time. 
This behavior is not evident in Bondi coordinates, as we show below.
Figure 1 shows the evolution of the radius $A$ in
a logaritmic scale.  Figure 2 displays the 
 profiles of the physical variables versus the time--like coordinate
 for the different comoving regions. 

 In order to explore the slow recovery of staticity at the
surface we force it to return to rest quickly after the
 emission
of energy; we prescribe its evolution instead of giving
the luminosity $L$. For such a prescription the surface equations change; now
we obtain the luminosity profile from a differential equation. 
We choose a radius evolving as
$$A(t)=\frac{(A_i-A_f)(e^{-t_d/\sigma} + 1)}{e^{(t-t_d)/\sigma+1}}+A_f\, ,$$
where $A_i$ is the initial radius, $A_f$ the final radius, $t_d$ the decay
time and $\sigma$ is the decay rate.
Figure 3 shows the luminosity profile and the prescribed radius.
Observe that it is necessary to absorb some quantity of energy (less than
the total emitted) to reach
a final constant radius. We confirmed the same behavior in Bondi
coordinates. It is interesting to note that in Ref. \cite{hjr80}
the interior profiles of the flux radiation (for the Schwarzschild
type model) have the same qualitative 
behavior as we show in Figure 3: It is clear the absorption of energy.

\subsection{Tolman VI--like model}
In this subsection we discuss the model obtained from Tolman's solution
VI \cite{hjr80}, \cite{t39}. Let us take
\begin{equation}
\tilde\rho=\frac{3g}{r^2},
\end{equation}
\begin{equation}
\tilde p=\frac{g}{r^2}\Bigg\{\frac{1-9Dr}{1-Dr}\Bigg\},
\end{equation}
where $g$ and $D$ are functions of $t$, which can be determined using
(\ref{eq:boun}). Thus, 
\begin{equation}
\tilde\rho=\frac{3(1-F)}{24\pi r^2},
\end{equation}
\begin{equation}
\tilde p=\frac{(1-F)}{24\pi r^2}\Bigg\{\frac{\psi_T-9\chi_T (r/a)}
{\psi_T-\chi_T(r/a)}\Bigg\},
\end{equation}
where
$$\chi_T=(3\Omega^2-1)(1-F)+6E(1+\Omega),$$

$$\psi_T=3(\Omega^2-3)(1-F)+6E(1+\Omega).$$

Using (\ref{eq:mass}) and (\ref{eq:nu}) we obtain
\begin{equation}
m=m_a r/a,
\end{equation}
\begin{equation}
e^{\nu}=F\Bigg\{\Biggl(\frac{\psi_T-\chi_T(r/a)}{\psi_T-\chi_T}\Biggr)^2
(r/a)\Bigg\}^{4(1-F)/3F}.
\end{equation}

In this model, the LHS of (\ref{eq:TOV}) results in
\begin{eqnarray}
\tilde R(t)=\frac{(1-F)}{12\pi a^3(\psi_T-\chi_T)}\Bigg\{\frac{(\psi_T-9\chi_T)
\chi_T}{2(\psi_T-\chi_T)}-\psi_T+\frac{9\chi_T}{2}\Bigg\}+\nonumber \\
\nonumber \\
\frac{1}{16\pi a^3 F}
[(1-F)(1+\Omega^2)+2E(1+\Omega)]^2,
\end{eqnarray}
which lets us write the third equation at the surface as
\begin{eqnarray}
\dot\Omega=\frac{2}{F(1-F)}\Bigg\{\frac{\dot F^2A}{4F}+\dot F\dot A
+\frac{\dot FA}{4F}\Big\{\dot F-\frac{\dot A}{A}[(1-F)(\Omega^2+1)
\nonumber\\
\nonumber\\
+2E(1+\Omega)]\Big\}
-\Big\{\dot EF+4\pi A^2F^2\tilde R+\frac{F^2}{A}
[(1-F)\Omega^2
\nonumber \\ \nonumber \\+2E(1+\Omega)]\Big\}\Bigg\}. 
\end{eqnarray}
Again the system can be numerically integrated for a reasonable set of
initial conditions (with $L$  being a gaussian as in the 
Schwarzschild--like model,
radiating away $1/100$ of the initial mass), as for example,
$$  A(0)=6.67, \,\, F(0)=0.70, \,\, \Omega=-0.17.$$

Figure 4 shows the evolution of the radius of the sphere. Initially the
fluid sphere collapses, but later it bounces.
Figure 5
gives the evolution of the matter variables for different regions. Note,
for the matter velocity profiles,
that some inner zones continue contracting after the bounce of the outermost
ones.

\section{CONCLUDING REMARKS}
We have sought the dependence of the EVM upon the Bondi
 coordinates.
The idea that we can always construct dynamical solutions from static
ones seems a general method. In this paper we accomplish
successfully such a construction,
 at least, for the Schwarzschild coordinates. 
Some research is in
progress for more realistic luminosity profiles and dissipative transport
mechanisms, considering extended thermodynamics under the time relaxation
approximation.
We considered in this paper two simple and idealized 
models, not deprived of physical meaning at all. 
We may hope that they contain some of the essential features of 
gravitational collapse inasmuch as we have fed the models with
some observational data (initial velocity and total mass
radiated.) Also, we can hope that our toy models could serve as
test beds for the numerical relativity methods and codes. 

\section*{Acknowledgements}
We benefited from research support by the Consejo de Investigaci\'on
under Grant CI-5-1001-0774/96 of the Universidad de Oriente.  We
thank the Postgrado en F\'\i sica program of the Universidad de 
Oriente for the support received and to Luis Lehner
for his valuable comments.

\begin{figure}
\centerline{\includegraphics[width=30pc]{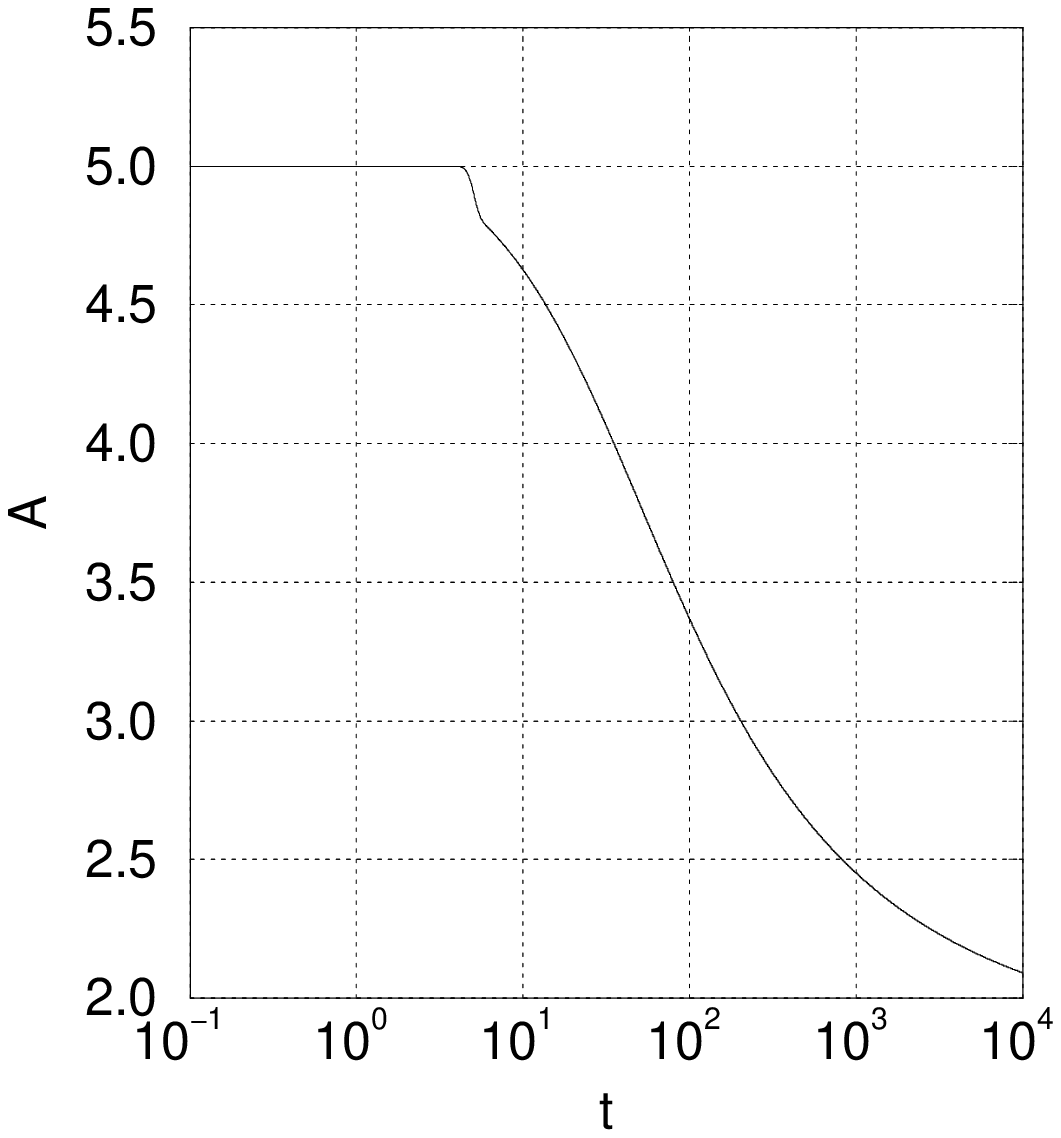}}
\caption{Evolution of the radius for the Schwarzschild type model.}
\label{fig:radiusS}
\end{figure}

\begin{figure}
\centerline{\includegraphics[width=30pc]{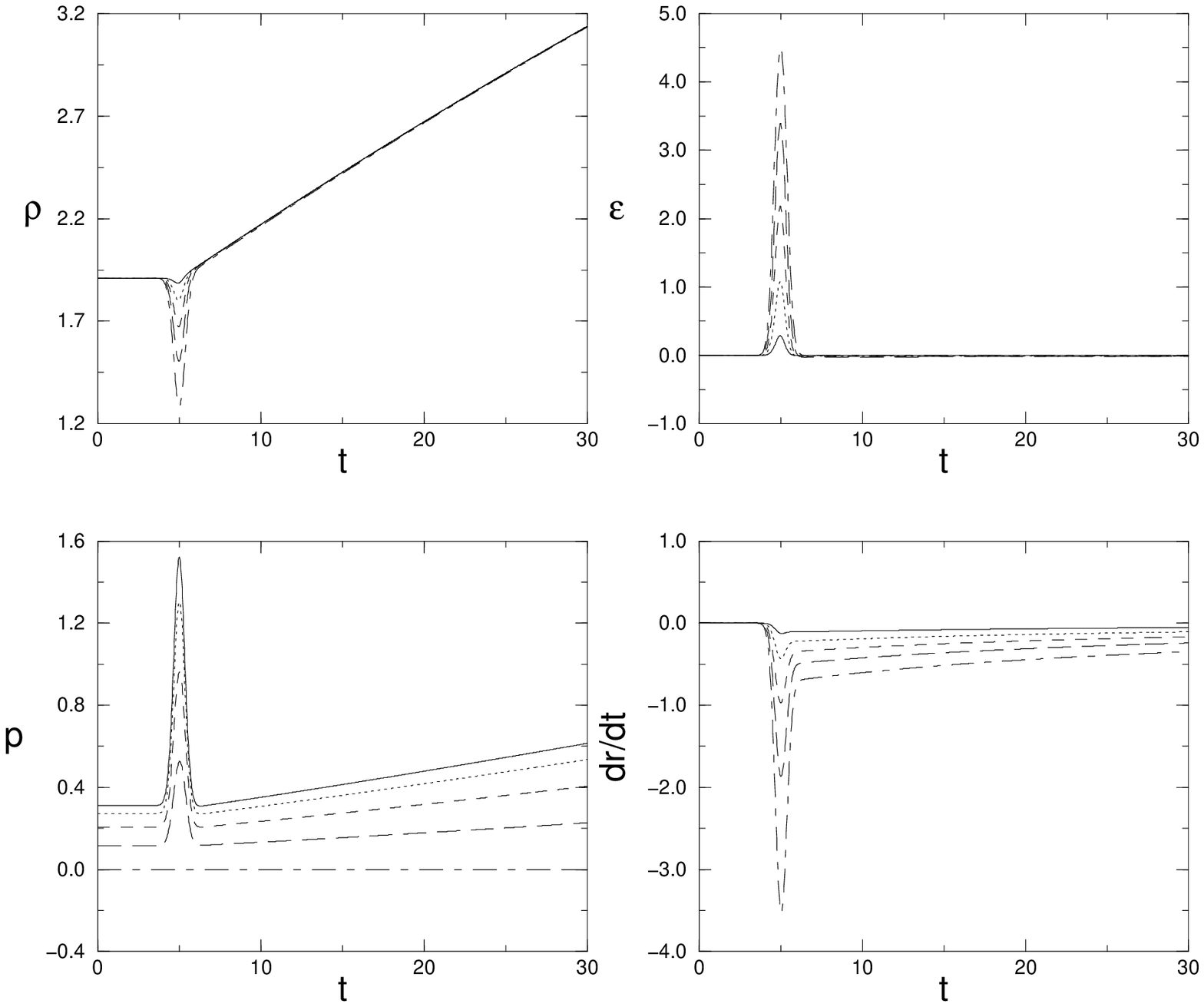}}
\caption{Density (multiplied by
$10^{3}$), energy density flux (multiplied by
$10^{4}$), pressure (multiplied by
$10^{3}$) and matter velocity (multiplied by $10$)
 for the Schwarzschild type model
 as a function of the time--like coordinate
 and different pieces of the material:
$0.2$ (solid line);
$0.4$ (dotted line); $0.6$ (small--dashed line);
 $0.8$ (dashed line); $1.0$ (dot--dashed line.)
}
\label{fig:S}
\end{figure}

\begin{figure}
\centerline{\includegraphics[width=25pc]{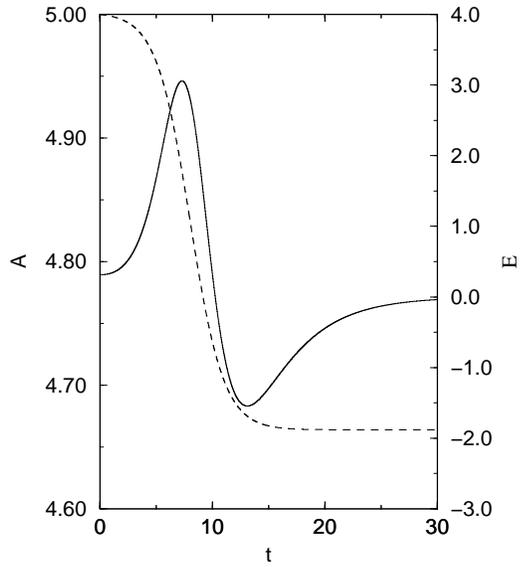}}
\centerline{\includegraphics[width=25pc]{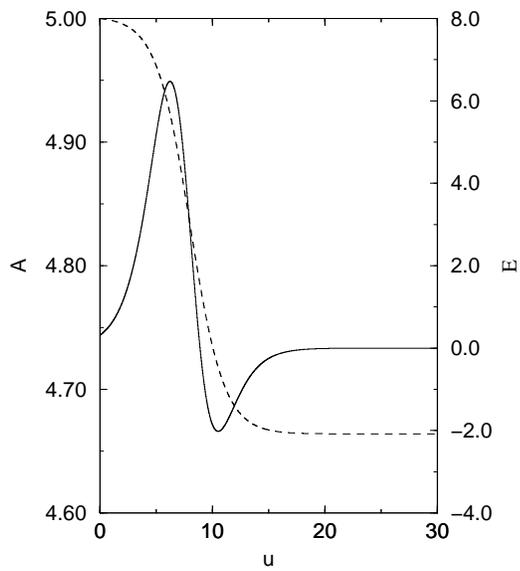}}
\caption{Evolution of the enforced radius $A$ (dashed line)
 and the resulting luminosity
$E$ (solid line multiplied by $10^2$) for the Schwarzschild
 type model in Schwarschild and Bondi coordinates.}
\label{fig:radius_B_SC}
\end{figure}
\begin{figure}
\centerline{\includegraphics[width=30pc]{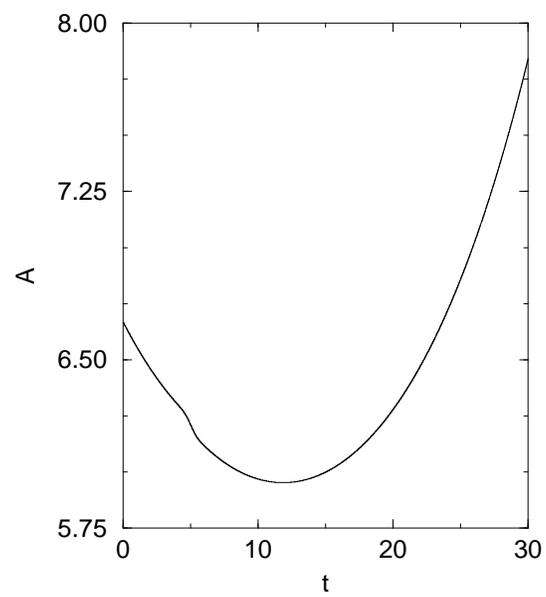}}
\caption{Evolution of the radius for the Tolman VI type model.}
\label{fig:radiusTVI}
\end{figure}

\begin{figure}
\centerline
{\includegraphics[width=30pc]{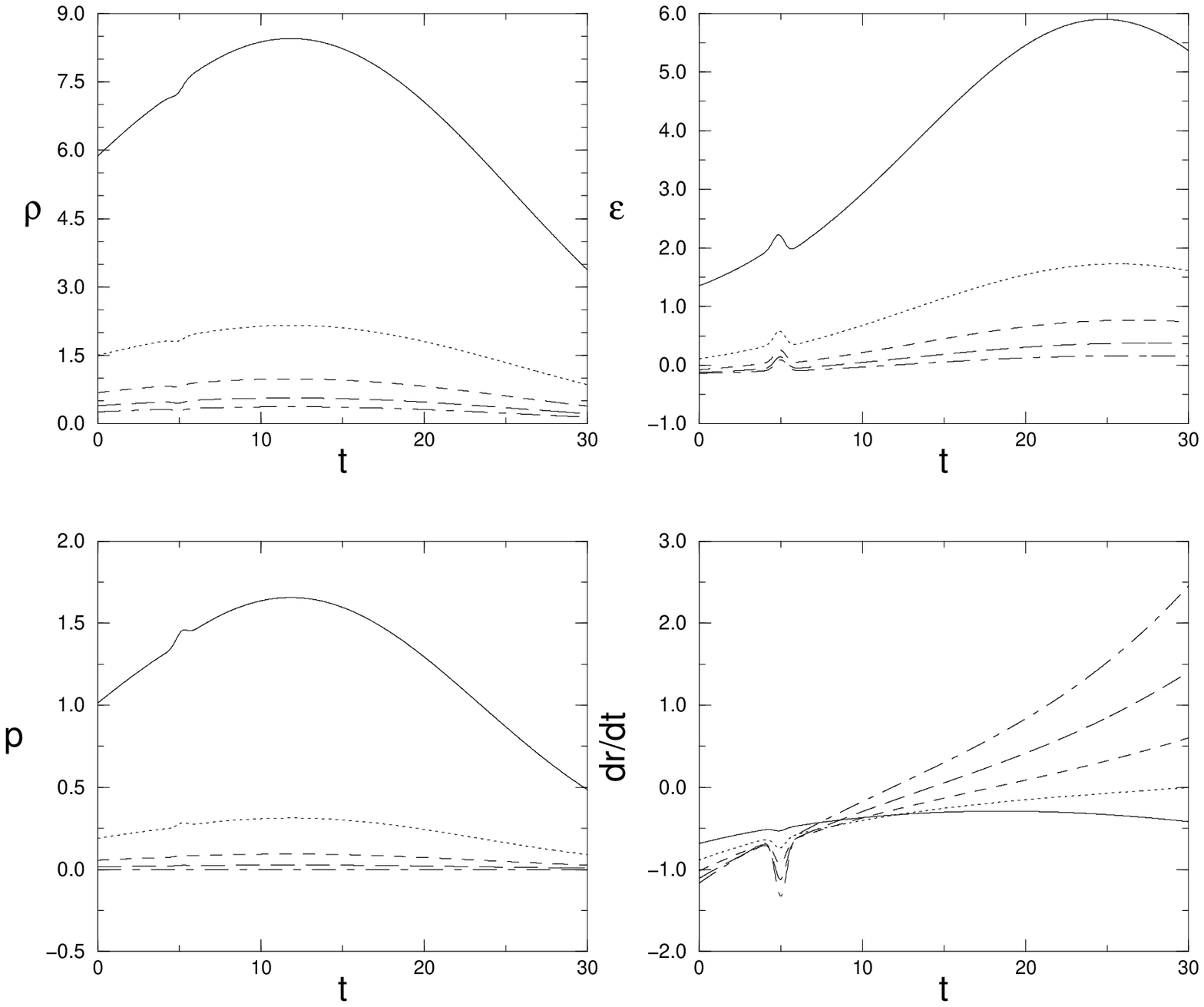}}
\caption{Density (multiplied by
$10^{3}$), energy density flux (multiplied by
$10^{4}$), pressure (multiplied by
$10^{3}$) and matter velocity (multiplied by $10$)
for the Tolman VI type model
 as a function of the time--like coordinate
and different pieces of the material:
$r/a=0.2$ (solid line);
$0.4$ (dotted line); $0.6$ (small--dashed line);
 $0.8$ (dashed line); $1.0$ (dot--dashed line.)
}\label{fig:TVI}
\end{figure}
\end{document}